\documentclass[aps,pra,twocolumn,noshowpacs]{revtex4}
\usepackage{graphicx,epsfig}
\usepackage{amsbsy}
\usepackage{amsmath}

\DeclareMathAlphabet{\mathvec}{OT1}{cmr}{bx}{sl}
\newcommand{\figref}[1]{\mbox{Fig.~\ref{#1}}}

\date{\today}

\begin{document}
\title{Non-equilibrium dynamics of an unstable quantum pendulum}
\author{C.S. Gerving, T.M. Hoang, B.J. Land, M. Anquez, C.D. Hamley,  and M.S. Chapman}

\affiliation{School of Physics, Georgia Institute of Technology,
  Atlanta, GA 30332-0430}

\maketitle

{\bf

A pendulum prepared perfectly inverted and motionless is a prototype of unstable equilibria and corresponds to an unstable hyperbolic fixed point in the dynamical phase space.  Unstable fixed points are central to understanding Hamiltonian chaos in classical systems \cite{Tabor}. In many-body quantum systems, mean-field approximations fail in the vicinity of unstable fixed points and lead to dynamics driven by quantum fluctuations \cite{Sachdev01, Dziarmaga10}. Here, we measure the non-equilibrium dynamics of a many-body quantum pendulum initialized to a hyperbolic fixed point of the phase space. The experiment uses a spin-1 Bose condensate \cite{Ho98, Ohmi98, Stenger99}, which exhibits Josephson dynamics in the spin populations that correspond in the mean-field limit to motion of a non-rigid mechanical pendulum \cite{Smerzi97,Zhang05}. The condensate is initialized to a minimum uncertainty spin state, and quantum fluctuations lead to non-linear spin evolution along a separatrix and non-Gaussian probability distributions that are measured to be in good agreement with exact quantum calculations up to 0.25~s. At longer times, atomic loss due to the finite lifetime of the condensate leads to larger spin oscillation amplitudes compared to no loss case as orbits depart from the separatrix.  This demonstrates how decoherence of a many-body system can result in more apparent coherent behaviour. This experiment provides new avenues for studying macroscopic spin systems in the quantum limit and for investigations of important topics in non-equilibrium quantum dynamics \cite{Polkovnikov11}.
}

A pendulum initialized to a hyperbolic fixed point is metastable in the classical limit. Phase orbits passing close to these points have exponentially diverging periods, and the orbits passing exactly through these points form a separatrix between librational and rotational motion of the pendulum with an infinite period. If the pendulum is prepared perfectly in this orientation, the classical equations of motion predict that it will not evolve. In reality, even if perfect preparation was possible, thermal fluctuations of the pendulum would perturb the pendulum from the metastable orientation and lead to oscillation.  Even at zero temperature, unavoidable quantum fluctuations would lead to evolution \cite{Cook86,Leibscher09}.  Although mechanical pendulums operating at the quantum limit are currently unavailable in the lab, it is possible to study quantum many-body systems that have similar dynamical behavior \cite{Albiez05, Chang05, Levy07}.

The focus of this work is spin-1 atomic Bose condensates \cite{Ho98, Ohmi98, Stenger99} with ferromagnetic interactions tightly confined in optical traps such that spin domain formation is energetically suppressed.  In this case, the non-trivial dynamical evolution of the system occurs only in the internal spin variables, and the mean-field dynamics of the system can be described by a non-rigid pendulum similar to the two site Bose-Hubbard model \cite{Smerzi97,Zhang05}.  The system is fully integrable in both the quantum \cite{Law98} and classical \cite{Pu99,Zhang05} limits, and exhibits a rich array of non-linear phenomena including Hamiltonian monodromy \cite{Lamacraft11}. Furthermore, the condensate features a tunable Hamiltonian with a quantum phase transition that permits quenching of the condensate to highly-excited spin states. Together, these provide unique capabilities to explore non-equilibrium quantum dynamics that are not captured by mean-field approaches and can be solved exactly with Schr\"{o}dinger's equation.

In these experiments, we study the evolution of a quenched spin-1 condensate prepared in a metastable state corresponding to a hyperbolic fixed point in the spin-nematic phase space that ultimately evolves far beyond the perturbative limit.
The quantum solution of the problem at zero magnetic field yields intricate spin-mixing dynamics that exhibit non-linear quantum revivals \cite{Law98} and a \emph{quantum carpet} of highly non-Gaussian fluctutations \cite{Diener06}. At finite fields, the dynamics are similar \cite{Chen09,Heinze10}, although they occur on a time-scale favorable for experimental observation. In both cases, the evolution occurs along a separatrix of the phase space and is driven by quantum fluctuations that are absent from the mean-field theory solutions \cite{Pu99,Heinze10}.

The equilibrium states, domain formation and spin dynamics of spinor condensates have been studied in many experiments \cite{Stenger99,Schmaljohann04,Chang04,Chang05,Kronjager06,Black07,Liu09,Leslie09,Klempt10,Bookjans11b,Gross11,Lucke11,Hamley12}.  In particular, observation of coherent spin oscillations have confirmed the mean-field pendulum model for small condensates  \cite{Chang05,Kronjager06,Black07}. Spin evolution has been previously observed from metastable spin states in many experiments \cite{Stenger99,Chang04,Klempt10,Bookjans11b,Gross11,Lucke11,Hamley12}, however, the experiments have not yet demonstrated spin dynamics in agreement with quantum calculations, except in the perturbative, low-depletion limit at very short times (where a Bogoliubov expansion around the mean field can be used) \cite{Leslie09,Klempt10,Hamley12} or for conditions where the mean-field approach suffices.  Here, by using low-noise atom detection techniques and careful state preparation, we are able to observe quantum spin dynamics that agree well with quantum calculations and demonstrate a rich array of non-Gaussian fluctuations.

We begin by discussing the exact quantum model for spin-1 condensate small enough to be described by a single domain.   The quantum states of the system can be described in a Fock basis, $|N_1,N_0,N_{-1}\rangle$, where $N_i$ are the number of atoms in the three spin-1 Zeeman states. The spin dynamics, including the effects of a magnetic field, are governed by the interaction Hamiltonian \cite{Law98,Chen09,Heinze10}:
\begin{eqnarray}
\label{Hamilton}
	\mathcal{H} &=& \lambda [(
	\hat{N}_{1} - \hat{N}_{-1})^2
	+ (2  \hat{N}_{0}-1) (\hat{N}_{1}+ \hat{N}_{-1})   \nonumber\\
	& &~~+ 2 \hat{a}_{0}^\dag \hat{a}_{0}^\dag \hat{a}_{1} \hat{a}_{-1}
	+ 2 \hat{a}_{1}^\dag \hat{a}_{-1}^\dag \hat{a}_{0} \hat{a}_{0}] \nonumber\\
	& &~~+q(\hat{N}_{1}+ \hat{N}_{-1}).
\end{eqnarray}

\noindent Here, $\hat{a}_{i} $  are the bosonic annihilation operators for the three spin states and $\hat{N}_{i}=\hat{a}_{i}^\dag \hat{a}_{i}$.  $\lambda$ and $q (\propto B^2 )$ characterize the inter-spin and Zeeman energies, respectively.  The spin-dependent binary collisions restrict the dynamical evolution to states that conserve both the total number of atoms $N=\sum_i N_i$ and the projection of angular momentum along the quantization axis $M=N_1-N_{-1}$. Starting from the initial state $|0,N,0\rangle$, consisting of all $N$ atoms in the $m_f = 0 $ state, the evolution is constrained to final states of the form $\sum_p c_p |p,N-2p,p\rangle$.  Hence, the  solution to the quantum many-body problem is fully enumerated by the time-dependence of the Fock state amplitudes, $c_p(t)$.

\begin{figure}[t!]
    \includegraphics{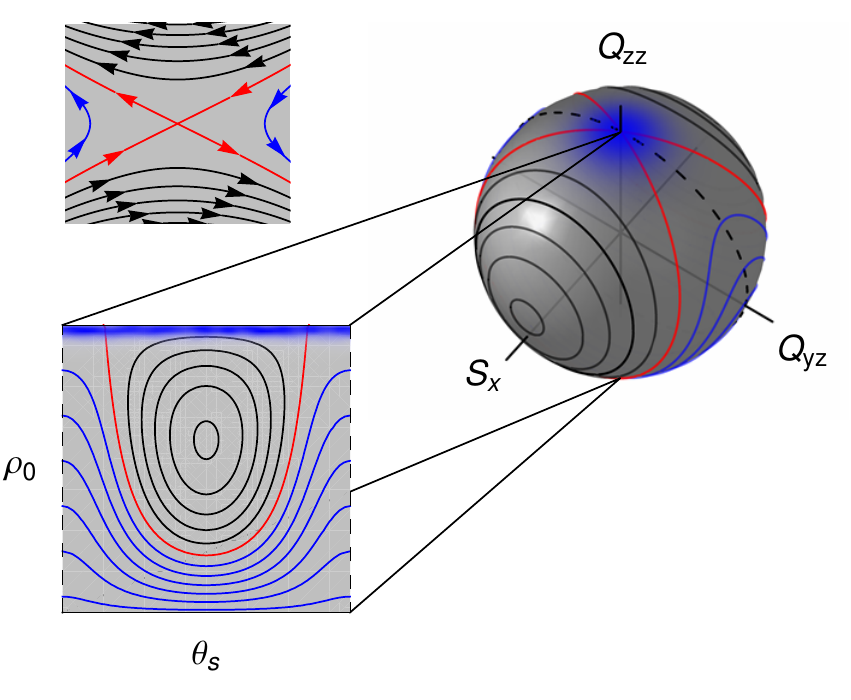}
\caption{\textbf{Phase Space} The spin state immediately following the quench is depicted on two relevant phase spaces of the spin-1 system: the $\{\rho_0,\theta_s\}$ phase space (bottom-left) and the  $\{S_x,Q_{yz},Q_{zz}\}$ spin-nematic Bloch sphere (right). A zoom-in of the hyperbolic fixed point at the pole is shown (top-left) with arrows indicating the  orbit directions.  The $\rho_0,\theta_s$ phase space represents a Mercator projection of the $\{S_x,Q_{yz},Q_{zz}\}$ sub-space.}
\label{PhaseSpace}
\end{figure}

The semi-classical dynamics of the system take the form of a non-rigid pendulum \cite{Zhang05}.
Mean field states of a spin-1 condensates can be written as $\psi=(\zeta_{+1},\zeta_0,\zeta_{-1})^T$ where $\zeta_i=\sqrt{\rho_i} e^{i \theta_{i}}$,  and $\rho_i =|\zeta_i|^2 =N_i/N$ are the fractional spin populations.
The conservation of magnetization $m=(N_1-N_{-1})/N$ constrains the populations $\rho_{\pm1}=(1-\rho_0 \pm m)/2 $, and for the $m=0$ case that is relevant for these experiments, the spin dynamics are determined by the Hamiltonian:
\begin{equation}
	\mathcal{H} = \lambda' x^2-\lambda'(1-x^2) \text{cos} \theta_s-q x
\label{mf}
\end{equation}

\noindent
Here,  $x=(\rho_0-1/2)/2$ and $\theta_s = \theta_{+1}+\theta_{-1}-2\theta_0$ are canonically conjugate variables and $\lambda' = 2N \lambda$. This Hamiltonian has the form of a classical non-rigid pendulum and is similar to the double-well Bose-Hubbard model that has been used to study Josephson effects in condensates.  The Hamiltonian can also be written using a phase space of the spin vector $S_i$ and nematic (quadrupole) tensor $Q_{ij}$ matrix operators for the spin-1 system:  $\mathcal{H} = \lambda' \sum_i S_i^2 +  q Q_{zz}/2$.  The phase spaces for both of these forms are shown in \figref{PhaseSpace}, where it is clear that the $\rho_0,\theta_s$ phase space corresponds to a projection of the spin-nematic phase space.

The initial state of the system following the quench, $|0,N,0\rangle$, is indicated in the different phase spaces in \figref{PhaseSpace}  using quasi-probability distributions of the initial state determined from the quantum uncertainties  \cite{Hamley12}.  In the spin-nematic space, the state corresponds to a minimum uncertainty state centered at the pole.  The pole is a hyperbolic fixed point lying at the intersection of the separatrix that separates the librational and rotational orbits of the system.  In the projected $\rho_0,\theta_s$ phase space, the distribution in $\rho_0$ is tightly packed at the top of the phase space with random spinor phase.  In the absence of quantum fluctuations, the state initialized the hyperbolic fixed point is non-evolving.  However, quantum fluctuations populate a family of orbits that straddle the fixed point, and subsequent evolution leads to phase flow along the unstable manifolds of the separatrix.  In the short term, this creates squeezed states with negligible change in $\rho_0$ \cite{Hamley12}.  For longer times, the system evolves along the separatrix, which forms a closed homoclinic orbit in the spin-nematic space.

\begin{figure}[t]
	\begin{center}
		\hspace{0.45in} \includegraphics{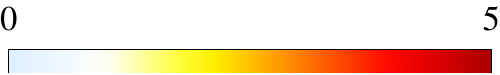} \\
    \includegraphics{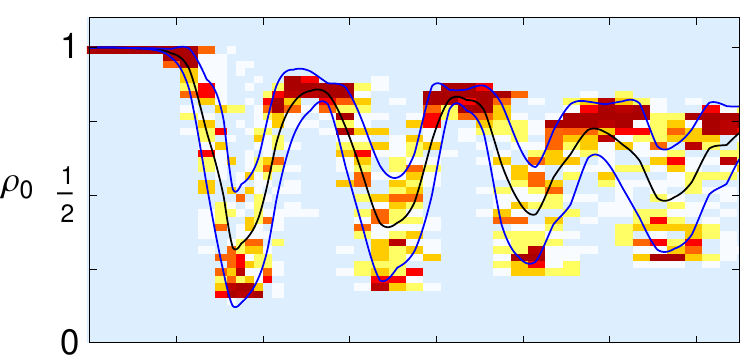} \\
    \includegraphics{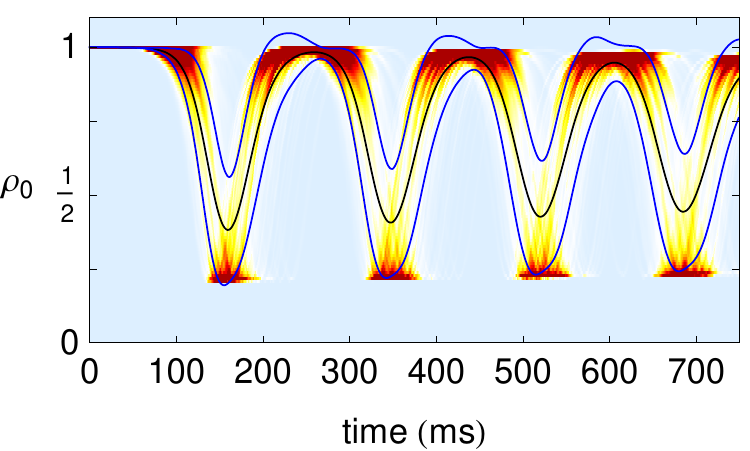}
	\end{center}
\caption{\textbf{Time evolution of spin populations.} Probability density of the fractional population of the condensate in the $m_f=0$ state, $\rho_0$, as a function of time.  The curves show the mean, $\bar{\rho}_0$ (black line) and $\pm$ the standard deviation, $\sigma$ (blue lines). \textbf{(a)} Experimental data showing the results of 50 runs at each evolution time placed into 40 bins.  The  mean and standard deviation curves have been smoothed using a cubic spline.  \textbf{(b)} Quantum calculation using the initial atom number, magnetic field ramp, and atom loss rate measured in the experiment.  The Fock state probabilities $|c_p|^2$, placed into 100 bins, are plotted.
}
	\label{Raw}
\end{figure}

\begin{figure}[p!]
	\begin{center}
	\begin{tabular}{ll}
		\textbf{\textsf{a}} &
		\textbf{\textsf{b}} \\
    \includegraphics[scale=1]{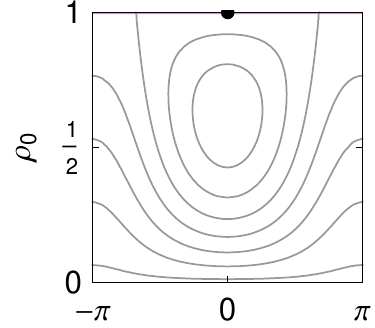} &
    \includegraphics[scale=1]{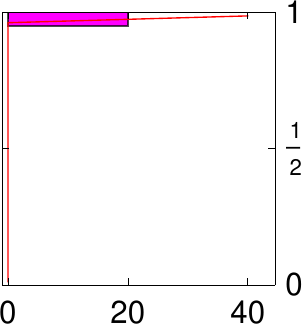} \\
    \includegraphics[scale=1]{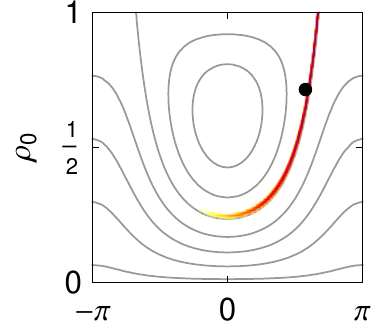} &
    \includegraphics[scale=1]{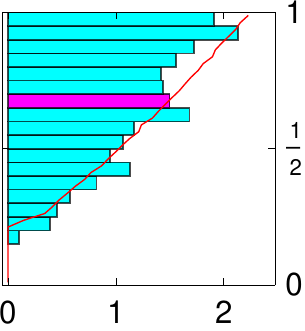} \\
    \includegraphics[scale=1]{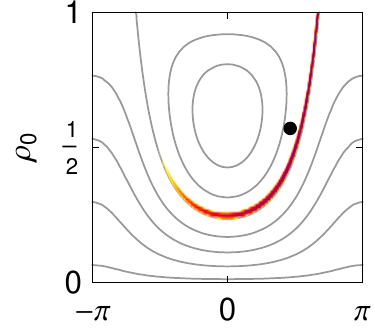} &
    \includegraphics[scale=1]{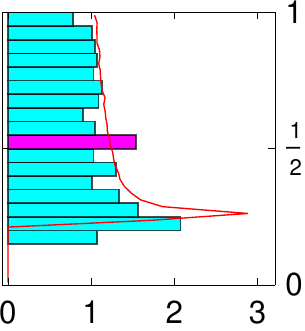} \\
    \includegraphics[scale=1]{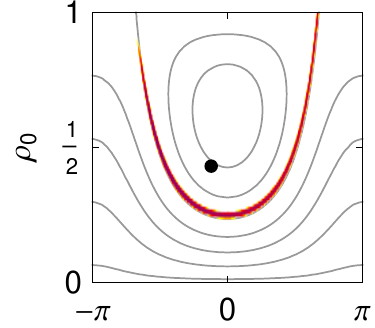} &
    \includegraphics[scale=1]{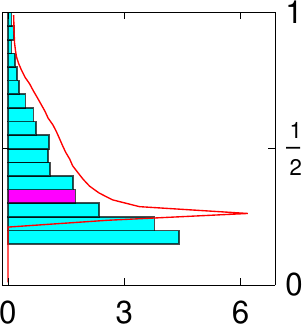} \\
    \includegraphics[scale=1]{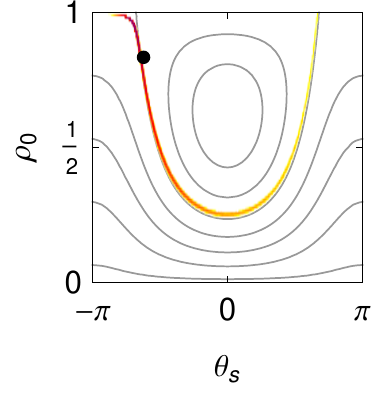} &
    \includegraphics[scale=1]{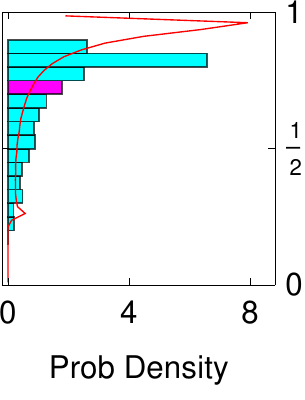}
    \end{tabular} \\
   \end{center}
\caption{\textbf{Full probability distributions of $\rho_0$.} Evolution on the semi-classical phase space (left column) and histograms of the measured spin population, $\rho_0$ (right) for different evolution times after the quench, 15~ms, 130~ ms, 140~ms, 170~ms, and 240~ms. \textbf{a. }The simulations use the semi-classical equations of motion together with a quasi-probability distribution for the initial state.  The mean value for $\rho_0$ and $\theta_s$ are indicated with a black dot.  \textbf{b. }The histogram bars for each evolution time depict the measured probability density of $\rho_0$ for over 900 experimental runs, and the red line represents the simulation.  The purple bar in each histogram represents the bin in which the mean of $\rho_0$ is located.
   }
\label{PDF}
\end{figure}

We now turn to the experimental results. The experiment begins with a rubidium-87 condensate containing $4 \times 10^4$ atoms, initialized in the $f=1,m_f=0$ hyperfine state and held in a high magnetic field.  The condensate is rapidly quenched by lowering the field, and the spin populations are measured for different evolution times. The experiment is repeated many times in order to acquire sufficient statistics to determine the full probability distributions of the populations. The main results of the paper are shown in \figref{Raw}, which shows the measured probability density of $\rho_0 = N_0/N$ versus evolution time, which is effectively a determination of the probabilities $|c_p|^2$.  The experimental results are compared with a quantum calculation using a spinor energy, $2\lambda N = -2 \pi \hbar \times 7.5~\mathrm{Hz}$, chosen to match the population dynamics. Both the experiment and quantum solutions exhibit population evolution that is in good overall agreement.  In particular, both exhibit a long pause (80~ms) before any population evolution is apparent.  After this pause, the spin population executes a regular damped oscillation. Population evolution from the metastable state is exponentially sensitive to initial population in the $m_F= \pm1$ states \cite{Klempt10}. At the earliest evolution time studied (15 ms), the total population in these states is measured to be $<30$ atoms which represents an upper bound limited by atom detection noise \cite{Hamley12}. Initial populations at this level effect the duration of the initial pause and first oscillation minimum, but not the overall character of the evolution \cite{Diener06} (see Supplemental Information).  For evolution times beyond  $>250$~ms, it is necessary to include in the theory the effects of atomic loss due to the lifetime of the condensate $\tau=1.8$~s, which is discussed in more detail below.

It is clear that the mean and standard deviation are insufficient to fully characterize the distribution of $\rho_0$  for both the experiment and theory, since for much of the evolution the mean does not pass through the highest probability density, and the asymmetry indicates a significant skew in the distribution.  This point is reinforced in \figref{PDF}, which shows the full probability distributions for several evolution times, along with the theoretical predictions. The highly non-Gaussian nature of the distributions provide compelling evidence of the quantum nature of the spin dynamics.
The physical origin of these non-Gaussian fluctuations is dispersion of neighboring orbits about the separatrix.  Immediately following the quench, the distribution in $\rho_0$ is tightly packed at the top of the phase space with random spinor phase.  This state corresponds to a minimum uncertainty state of the spin-nematic subspace shown in \figref{PhaseSpace} \cite{Hamley12}. As evolution proceeds, the phase, $\theta_s$, converges towards the separatrix separating the librational and rotational trajectories, and the population starts to evolve along it. The separatrix has a divergent period \cite{Zhang05}, and so the states disperse significantly due to the different evolution rates of nearby energy contours. It is this dispersion, together with the shape of the orbit, that gives rise to the highly non-Gaussian probability distributions.

\begin{figure*}[t!]
	\begin{center}
	\begin{tabular}{cp{0.15in}cp{0.15in}c}
		\includegraphics[scale=1]{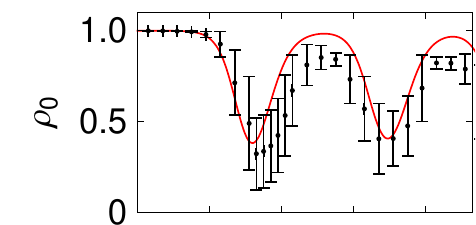} & &
		\includegraphics[scale=1]{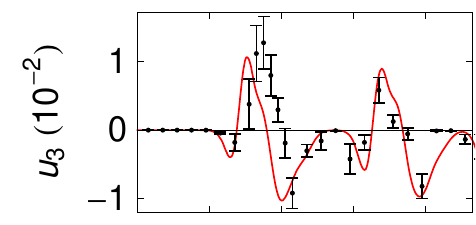} & &
		\includegraphics[scale=1]{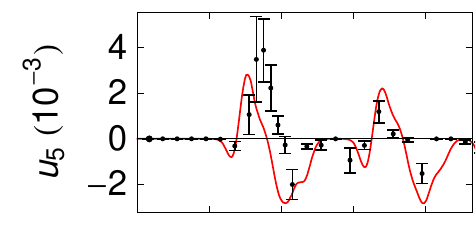} \\
		\includegraphics[scale=1]{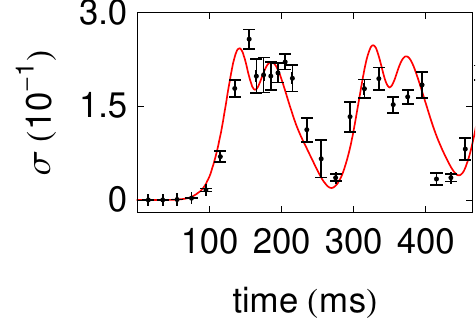} & &
		\includegraphics[scale=1]{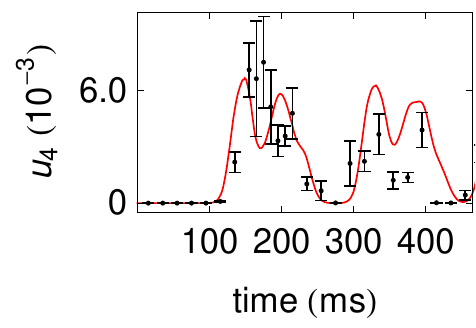} & &
		\includegraphics[scale=1]{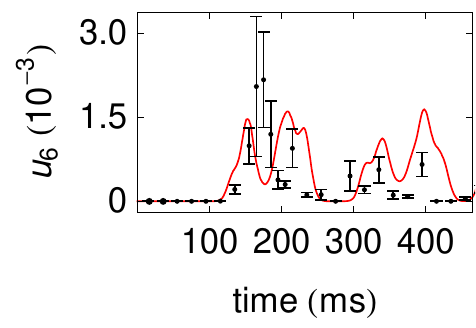}
	\end{tabular}
	\end{center}
\caption{\textbf{Evolution of $\rho_0$ central moments.} The mean value of $\rho_0$, the standard deviation $\sigma$, and the third through sixth central moments, $u_k = \langle (\rho - \bar{\rho})^k \rangle$, $k=3-6$.  The odd moments are on top, and the even moments are on the bottom.  In each plot, the black markers represent the results of 50 experimental runs, and the error bars are estimated using a bootstrap method for the third moment and higher.  The red curves are the prediction of the quantum simulation including loss.}
\label{Moments}
\end{figure*}

\begin{figure}[t!]
		\includegraphics[scale=1]{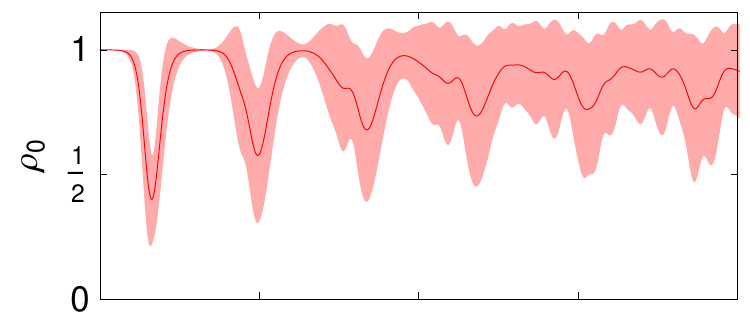} \\
		\includegraphics[scale=1]{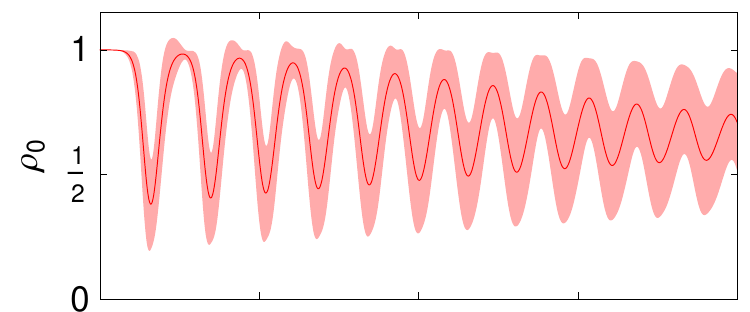} \\
		\includegraphics[scale=1]{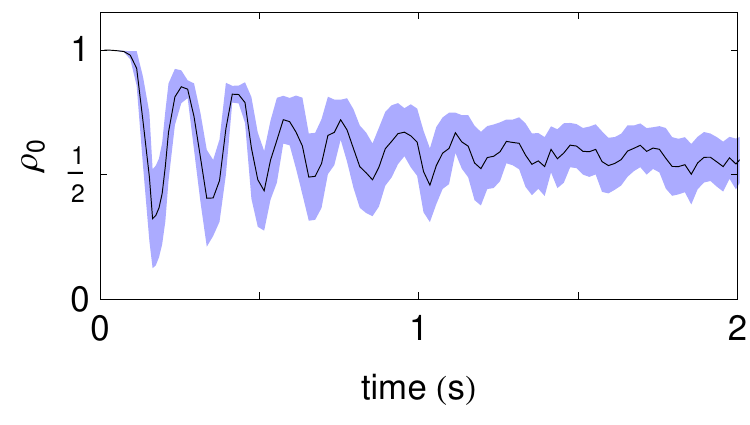}
\caption{\textbf{Long-term evolution of $\rho_0$.} The top graph shows the quantum calculation without loss. The middle graphs shows the calculation including the effects of uncorrelated atom loss.  The bottom graph shows the measured data.  In each plot, the mean value is shown as a solid line, and the shaded envelopes correspond to the  standard deviation}	
	\label{Moments2}
\end{figure}

In order to characterize the non-Gaussian distribution, we determine several central moments, $u_k = \langle (\rho - \bar{\rho})^k \rangle$ from the data.  The first six central moments are shown in \figref{Moments} compared with the quantum simulation. Overall the measured moments are in good agreement with the predicted moments from the simulation.  The population revival in the second oscillation predicted from the simulation is clearly seen in the first four moments, but is less obvious in higher moments.

We now turn to a discussion of the role of atomic loss in the dynamical evolution. The lifetime of the condensate $\tau=1.8$~s is only a factor of 10 larger than the spin evolution timescale ($\sim$150~ms), hence one expects that loss plays an important role in the dynamics.  We explore this question in \figref{Moments2} where we compare quantum calculations without loss, quantum calculations including uncorrelated loss and the experimental data. Uncorrelated atom loss is incorporated into the calculation using quantum Monte Carlo (QMC) techniques with the collapse operators $C_i = \sqrt{1/\tau}\, \hat{a}_i$.  The loss causes the overall magnetization $M$ to execute a random walk with a restoring tendency towards $M=0$ and decreases the spinor dynamical rate, which scales as  $\lambda \propto N^{-3/5}$.  (Supplementary Information)

For the first 250~ms of evolution corresponding to the first spin oscillation, the effects of loss are not discernable between the two calculations, and the experimental data are in good agreement with both. Beyond 250~ms, there are significant differences between the two quantum calculations. The spin population of the calculation without loss nearly returns to the initial value and then experiences a long pause followed by complex multi-frequency oscillations. The calculation with loss however exhibits steady oscillations with one dominant frequency and a slowly decreasing amplitude centered on the ground state populations.  In the semi-classical picture, the apparent damping of the calculation without loss derives from the dispersion about the separatrix in \figref{PDF}.  The effect of loss is to eventually move the orbits away from the separatrix, which turns off this dispersion and leads to more regular oscillations.

While the inclusion of loss into the model makes a significant improvement in the agreement of long term dynamics ($>250$~ms) with the experimental results, it is clear that this simple loss model is inadequate to fully replicate the measurements at longer time scales. While the experimental data and the simulations with loss are qualitatively similar, there is clearly more dissipation in the experiment as the amplitude of the oscillations damp more quickly and the standard deviation decreases.  In future work, we intend to further investigate the damping of the spin dynamics and its connection to thermalization of isolated quantum systems subject to loss. Similar investigations are on-going using 1-D condensate systems \cite{Kinoshita06,Hofferberth07,Hofferberth08,Trotzky12}, and it will be interesting to explore the similarities and differences in these completely different systems.  Finally, we believe that our results point the way to a host of fascinating explorations of out-of-equilibrium quantum spin systems \cite{Dziarmaga10,Polkovnikov11}.

\noindent \textbf{Methods}

We prepare a  condensate of $N= 38,500\pm500$ $^{87}$Rb atoms in the  $|f=1,m_f=0 \rangle$ hyperfine state in a high magnetic field ($2~\mathrm{G}$). The condensate is tightly confined in an optical dipole trap with trap frequencies of $  250~\mathrm{Hz}$.  To initiate dynamical evolution, the condensate is quenched below the quantum critical point  by lowering the magnetic field to a value $210~\mathrm{mG}$ and then allowed to freely evolve for a set time.  The trap is then turned off and a Stern-Gerlach field is applied to separate the $m_f$ components during 22~ms time-of-flight expansion. The atoms are probed  for $400~\mu\mathrm{s}$ with three pairs of orthogonal laser beams, and the resulting fluorescence signal is collected by a CCD camera with $>90\%$ quantum efficiency.

\bibliography{QPref}
\bibliographystyle{nature}

\end{document}


\title{Non-equilibrium dynamics of an unstable quantum pendulum:  Supplementary Information}
\author{C.S. Gerving, T.M. Hoang, B.J. Land, M. Anquez, C.D. Hamley,  and M.S. Chapman}
\affiliation{School of Physics, Georgia Institute of Technology,
  Atlanta, GA 30332-0430}

\maketitle

In this Supplementary Information, we provide an overview of the calculation methods used to compare the experimental results to theory and discuss the effect of spin impurities in the initial state.
\section{Second Quantized Calculations}
\subsection{Exact quantum calculations}
The first calculation method is the exact second quantized form of the Hamiltonian.  The second quantized form operates on a Fock basis represented by $| N_1,N_0,N_{-1} \rangle$ where $N_i$ is the number of particles in the $m_f$ state given in the index.  The spin mixing Hamiltonian conserves both the total atom number $N=N_1+N_0+N_{-1}$ and the magnetization $M=N_1-N_{-1}$.
The Hamiltonian in this basis is given by:
\begin{eqnarray}
\label{Hamilton}
	H_{SMA} &=& \lambda [(
	\hat{N}_{1} - \hat{N}_{-1})^2
	+ (2  \hat{N}_{0}-1) (\hat{N}_{1}+ \hat{N}_{-1})   \nonumber\\
	& &~~+ 2 \hat{a}_{0}^\dag \hat{a}_{0}^\dag \hat{a}_{1} \hat{a}_{-1}
	+ 2 \hat{a}_{1}^\dag \hat{a}_{-1}^\dag \hat{a}_{0} \hat{a}_{0}] \nonumber\\
	& &~~+q(\hat{N}_{1}+ \hat{N}_{-1}).
\end{eqnarray}
Similar to Ref. \cite{Law98,Diener06,Mias08,Chen09,Heinze10}, we write the Hamiltonian in a Fock basis of $N$, $M$, and $k$, where $N_0=N-2k$.  $k$ represents the number of pairs of atoms that have evolved from the $m_f=0$ state via spin-mixing.
In matrix form operating on the vector of $k$ coefficients this can be written:
\begin{eqnarray}
\label{BTriDiagonal}
\lefteqn{\tilde{H}_{k,k'} =} \nonumber \\
& & \{ \lambda M^2 + 2 \lambda k(2(N-2k)-1) + q (2k+|M|) \} \delta_{k,k'} \nonumber \\
 &+& 2 \lambda \{ (k'+1) \sqrt{(N-2k')(N-2k'-1)} \delta_{k,k'+1} \nonumber \\
 &+& k'\sqrt{(N-2k'+1)(N-2k'+2)} \delta_{k,k'-1} \}.
\end{eqnarray}

\noindent where $\lambda = \frac{c_{2}}{2} \int d\textbf{r} \left|\phi (\textbf{r})\right|^{4}$, $q = \mu_B^2 B_z^2 / \left(\hbar^2 E_{HFS} \right)$, $c_{2}$ is the spin-dependent collisional interaction strength, and $E_{HFS}$ is the ground state hyperfine splitting.  $\lambda$ and $q$ characterize the inter-spin and Zeeman energies respectively. This produces a tridiagonal matrix which can be exactly solved or numerically integrated.  In our simulation we numerically integrate since the magnetic field and spinor energy hence $q$ and $\lambda$ vary in time.  The value for $q(t)$ is modeled from experimental measurements of the magnetic field using microwave spectroscopy.  $\lambda$ varies with the atom number as $N^{-3/5}$.\footnote{This scaling law is determined by integrating the Thomas-Fermi density profile of a condensate confined in a three dimensional harmonic trap.}  The simplest way to account for this in the dynamical simulations is to make $\lambda$ a function of time.  Since $N$ decays approximately exponentially with atom loss, the value for $\lambda(t)$ is estimated from population dynamics and then scaled according to the condensate lifetime.  This simple model captures much of the early dynamics, but does not produce as much damping as observed experimentally.

\subsection{Quantum Monte Carlo}


A more rigorous calculation of the effects of atom loss is to use a quantum Monte Carlo simulation.  The quantum Monte Carlo is implemented similar to Refs. \cite{PhysRevA.46.4363,carmichael1993open,Tan99}.
The atoms are assumed to be lost one at a time and the process of losing an atom effectively measures its $m_f$ state and so the collapse operators are simply related to the annihilation operators for the modes of the condensate.
The numerical integration of the $k$ coefficients is performed with an effective Hamiltonian,
\begin{eqnarray}
	H_{eff} &=&  H_{SMA} -\frac{\imath}{2} \sum_{i}{C_i^\dag C_i^{\textcolor{white}{\dag}}} \nonumber \\
	&=& H_{SMA} -\frac{\imath}{2\tau} \hat{N}
\end{eqnarray}
where $C_i = \sqrt{1/\tau} \,\hat{a}_i$ are the collapse operators for each mode ($i=-1,0,1$) and $\tau$ is the condensate lifetime.  During the time interval $\Delta t$ of the numerical integration each atom has a probability $e^{-\Delta t/\tau}$ of remaining.  For each atom a random number in the range 0 to 1 is generated to stochastically determine how many atoms to annihilate in each mode.  If this number is greater than $e^{-\Delta t/\tau}$, then the appropriate collapse operator is applied to the state vector.  The number of atoms for each mode is given by $\langle \hat{N}_i \rangle$.  After the collapse operators have been applied the $k$ coefficients are renormalized and the next step of the numerical integration is performed with updated values for $N$ and $M$. Results are obtained from the quantum Monte Carlo simulation by taking the average of quantum expectation values from many runs with the same initial conditions but a uniquely seeded sequence of random numbers to determine the annihilation probabilities. In effect, the results of the quantum Monte Carlo simulation are the average of many quantum trajectories.

At first glance the quantum Monte Carlo is a daunting task since in general it should be necessary to use a basis spanning every possible value of $N(t)$, $M(t)$, and $k(t)$ which scales as $N^3$.  However the action of the collapse operators shifts the state vector from $N(t)$ and $M(t)$ to $N(t+\Delta t)$ and $M(t+\Delta t)$ while modifying the $k$ coefficients in a well characterized way.  At any given step of the calculation there is only one value for $N$ and $M$.  So for any step of the calculation the basis is proportional to $N(t)$ and is completely described with the current values of $N$, $M$, and the complex coefficients for the $k$ index.

\section{Mean Field Calculations}
\subsection{Semi-classical simulations}
To make connections with the mean-field theory, we use a semi-class\-ical technique together with quasi-prob\-a\-bil\-i\-ty distributions (QPD) to regain the quantum statistics.  The mean-field order parameter is represented by a complex vector $(\zeta_1,\zeta_0,\zeta_{-1})^T$ where $\zeta_i$ represents the amplitude and phase of the classical field for the mode associated with the $m_f$ state given in the index.  A mean field analysis of the spin Hamiltonian \cite{Ho98,Ohmi98,Stenger99} produces the dynamical equations \cite{Pu99}:
\begin{subequations}
\begin{eqnarray}
	i \hbar \dot{\zeta}_1^{\textcolor{white}{*}} &=& E_1 \zeta_1^{\textcolor{white}{*}} +
		c [(\rho_1 + \rho_0 - \rho_{-1})\zeta_1^{\textcolor{white}{*}} + \zeta_0^2 \zeta_{-1}^*] \\
	i \hbar \dot{\zeta}_{0^{\textcolor{white}{*}}} &=& E_{0} \zeta_{0}^{\textcolor{white}{*}} +
		c [(\rho_{1} + \rho_{-1})\zeta_{0}^{\textcolor{white}{*}} + 2 \zeta_1^{\textcolor{white}{*}} \zeta_{-1}^{\textcolor{white}{*}} \zeta_{0}^*] \\
	i \hbar \dot{\zeta}_{-1}^{\textcolor{white}{*}} &=& E_{-1}^{\textcolor{white}{*}} \zeta_{-1} +
		c [(\rho_{-1} + \rho_0 - \rho_{1})\zeta_{-1}^{\textcolor{white}{*}} + \zeta_0^2 \zeta_{1}^*]~~~~
\end{eqnarray}
\end{subequations}
where $c = 2 N \lambda$, $\rho_i \equiv N_i/N = |\zeta_i|^2$ is the fractional population of the $m_f=i$ component, and $E_i$ is the magnetic field energy for each mode.

A convenient parameterization of the order parameter is given by
\begin{eqnarray}
	\zeta_1 &=& \sqrt{\frac{1-\rho_0+m}{2}} e^{\imath \frac{\theta_s+\theta_m}{2}} \nonumber \\
	\zeta_0 &=& \sqrt{\rho_0} \nonumber \\
	\zeta_{-1} &=& \sqrt{\frac{1-\rho_0-m}{2}} e^{\imath \frac{\theta_s-\theta_m}{2}} \nonumber \\
\end{eqnarray}
where $\theta_m = \theta_1-\theta_{-1}$ is the magnetization or Larmor precession phase, $\theta_s = \theta_1+\theta_{-1}-2\theta_0$ is the spinor phase,
and $m=(N_1-N_{-1})/N$ is the fractional magnetization.

Using this parameterization,
the dynamical equations reduce to \cite{Zhang05}:
\begin{eqnarray}
	\dot{\rho_0} &=& \frac{2 c}{\hbar} \rho_0 \sqrt{(1-\rho_0)^2 - m^2} \sin \theta_s \nonumber \\
    \dot{\theta_s} &=& -\frac{2 q}{\hbar} + \frac{2 c}{\hbar} \times \nonumber \\
    & &\left[ (1-2\rho_0) + \frac{(1-\rho_0)(1-2\rho_0)-m^2}
	{\sqrt{(1-\rho_0)^2 - m^2}} \cos \theta_s \right].
\end{eqnarray}

$\rho_0$ and $\theta_s$ are canonically conjugate variables and the dynamical phase space is defined by the corresponding spin energy functional of the condensate:
\begin{eqnarray}
	\mathcal{E} &=& \frac{c}{2} m^2
	+ c \rho_0 \left[ (1-\rho_0) + \sqrt{(1-\rho_0)^2 - m^2} \cos \theta_s \right] \nonumber \\
	& &+ p m + q (1-\rho_0).
\end{eqnarray}
$m$ and $\theta_m$ are also canonically conjugate variables.  However, $\theta_m$ is cyclic because it does not appear in the energy functional, and thus $m$ is conserved.

\subsubsection{SU(3)}

A spin-1 condensate can be described using the SU(3) symmetry group with the Cartesian dipole-quadrupole
decomposition of the Lie algebra su(3) as a basis.  There are three dipole (or angular momentum) operators $S_a$, and nine quadrupole operators $Q_{ab}$ which are moments of the quadrupole tensor ($\{a,b\} \in \{x,y,z\}$).  Only five of quadrupole operators are linearly independent since the quadrupole tensor is symmetric and traceless.
These operators are defined as \cite{Carusotto04,Di10}:
\begin{equation}
	S_a = - i \hbar \epsilon_{abc} b_b^\dag b_c^{\textcolor{white}{\dag}}
\end{equation}
\begin{equation}
	Q_{ab} = - b_a^\dag b_b^{\textcolor{white}{\dag}} - b_b^\dag b_a^{\textcolor{white}{\dag}}	+ \frac{2}{3} \delta_{ab} b_c^\dag b_c^{\textcolor{white}{\dag}}
\end{equation}
summing over repeated indices where
\begin{eqnarray}
b_x^\dag &=& \left(-a_{+1}^\dag+a_{-1}^\dag\right)/\sqrt{2} \nonumber \\
b_y^\dag &=& i\left(a_{+1}^\dag+a_{-1}^\dag\right)/\sqrt{2} \nonumber \\
b_z^\dag &=& a_0^\dag \nonumber
\end{eqnarray}

In Figure 3 of the main text we depict an SU(2) subspace of this SU(3) as $\{S_x, Q_{yz},Q_{zz}\}$.  Just as in \cite{Hamley12}, the exact subspace is  $\{S_x, Q_{yz},(Q_{zz}-Q_{yy})\}$.  However, our initial state of $m_f= 0$ makes the off-diagonal elements ($\propto\ \sqrt{N_1 N_{-1} }$) of $(Q_{zz}-Q_{yy})$ small and always average to zero over a Larmor precession cycle, so ignoring these terms $(Q_{zz}-Q_{yy})\approx \frac{3}{2} Q_{zz}$.  Additionally, $Q_{zz}$ is linear with $N_0$ ( $Q_{zz} = \frac{2}{3}(N-3N_0)$), making for an easy mapping to $\rho_0 \equiv N_0/N$.

For the initial state of all atoms in the $m_f=0$ level, $ |0,N,0\rangle$, we have the following expectation values for the variances: $\langle ( \Delta \hat{S}_x ) ^2 \rangle = \langle ( \Delta \hat{S}_y ) ^2 \rangle =\langle ( \Delta \hat{Q}_{xz} ) ^2 \rangle = \langle ( \Delta \hat{Q}_{yz} ) ^2 \rangle =N $

\subsubsection{Quasiprobability distributions}
It is well known that the perfect $m_f=0$ state does not evolve from these mean-field equations.  Some seeding is necessary.  This seeding comes from the quantum noise of the initial Fock state, $|0,N,0\rangle$.  In order to produce the necessary QPD, we note that the $|0,N,0\rangle$ Fock state is a minimum uncertainty coherent state with $S_x$, $Q_{yz}$, $S_y$, and $Q_{xz}$ all having an expectation value of zero and variance of $N$.  Here, $S_a$  ($\{a\} \in \{x,y,z\}$) are  the spin-1 dipole (or angular momentum) operators, and $Q_{ab}$ ($\{a,b\} \in \{x,y,z\}$) are  moments of the spin-1 quadrupole (nematic) tensor.  
Generating a sample of random numbers for each of these distributions, we can then find values for the initial $\zeta_1$, $\zeta_0$, and $\zeta_{-1}$ (or $\rho_0$ and $\theta_s$) to plug into the spinor dynamical equations for the semi-classical simulation \cite{Hamley12}.  The conversion equations are:
\begin{equation}
\label{ChiPlus}
	\tan \chi_+ = -\frac{S_y + Q_{yz}}{S_x + Q_{xz}} \nonumber
\end{equation}
\begin{equation}
\label{ChiMinus}
	\tan \chi_- = \frac{S_y - Q_{yz}}{S_x - Q_{xz}} \nonumber
\end{equation}
\begin{equation}
\label{Rho0}
	\rho_0 = \frac{1}{2} + \sqrt{\frac{1}{4} -
	\frac{1}{8}\left(\left( \frac{S_x + Q_{xz}}{\cos \chi_+} \right)^2 +
	\left( \frac{S_x - Q_{xz}}{\cos \chi_-} \right)^2 \right)} \nonumber
\end{equation}
\begin{equation}
\label{M}
	m = \rho_1 - \rho_{-1} = \frac{1}{8\rho_0}\left(\left( \frac{S_x + Q_{xz}}{\cos \chi_+} \right)^2 -	 \left( \frac{S_x - Q_{xz}}{\cos \chi_-} \right)^2 \right) \nonumber
\end{equation}
where $\chi_\pm = \theta_{\pm1} - \theta_0 = \frac{\theta_s \pm \theta_m}{2}$

Each of these sample wavefunctions is numerically integrated using the dynamical equations to yield $\rho_0(t)$ and $\theta_s(t)$, which are then plotted on the semi-classical phase space as in Fig.~2 in the main paper.

\section{Effect of Initial State Impurities}

\begin{figure}
\includegraphics{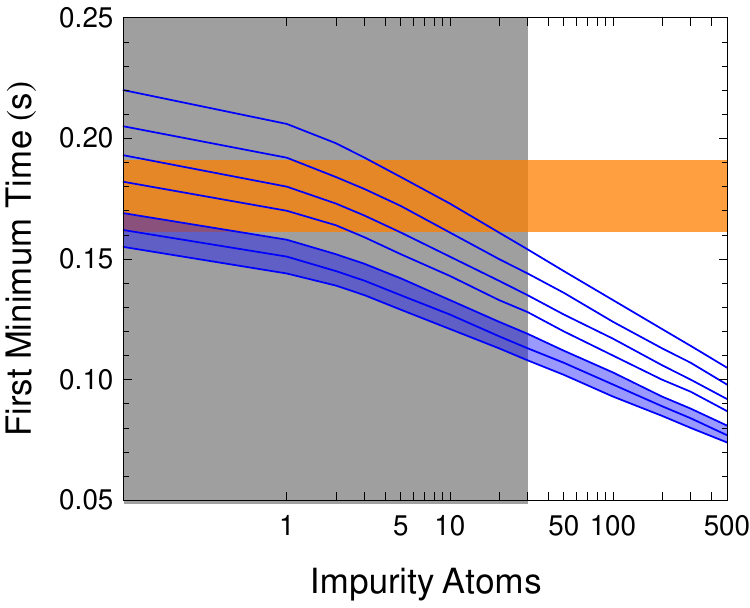}
\caption{{\bf Analysis of impurities.}  Upper bound from direct measure of impurities at 15~ms of evolution, gray shaded region.  Measurement of first minimum of $\rho_0$, orange shaded area.  Simulation with polluting atoms for a spinor dynamical rate of $c/(2 \pi \hbar)=$ 5.0, 5.5, 6.0, 6.5, 7.0, 7.5, and 8.0~Hz (blue lines top to bottom).  The values between 7-8~Hz are most consistent with the long term dynamics and are shaded blue.}
\label{Pollution}
\end{figure}

In this section, we discuss the effects of spin impurities in the initial state preparation, and the impurity limits determined by the experiment. The experiment uses condensates containing $4 \times 10^4$ atoms, initialized in the $f=1,m_f=0$ hyperfine state with measured impurities in the $m_f= \pm 1$ states (limited by the atom counting noise \cite{Hamley12}) below $<30$ atoms or 0.1\% of the total population.

Spin evolution from the metastable state is a parametric amplification process whose early-time dynamics are exponentially sensitive to initial population in the $m_F= \pm1$ states \cite{Klempt10}. Hence, any impurities in the initial state preparation will certainly effect the timescale of the initial pause and first oscillation minimum. Importantly though, as shown in \cite{Diener06}, the overall character of the evolution, including the intricate evolution of the quantum spin fluctuations, is robust to impurities even up to the few percent level, which is an order of magnitude larger than our measured bound.

In order to analyze the quantitative effect of impurities, we perform simulations with two types of impurities:  an initial non-zero magnetization and an initial non-zero number of pairs of $m_f=\pm1$ atoms.  The results of these calculations are nearly identical for the same number of impurity atoms with the non-zero magnetization results shown in Figure \ref{Pollution} for various levels of impurities and for a range of spinor dynamical rates determined from the long time evolution of the experiment (blue shaded region) as well as several other values (blue lines).  These are compared to experimental measurements in order to ascertain an upper bound on the impurities in the experiment.  The first time at which the atom populations are measured is $15~\mathrm{ms}$ after the beginning of the magnetic quench, which provides an upper bound on the impurities present at $t=0$.  The population in the $m_F = \pm1$ states at this time is $<30$ atoms, which is shown as the gray shaded region in Fig. \ref{Pollution}.   Also plotted in Fig. \ref{Pollution}  is the measured time that the $m_f=0$ population reaches a minimum value.  This time is exponentially sensitive to impurity atoms and, for both magnetization and pair impurities, reduces similarly for the same number of impurity atoms.  The experimental measurement of this time plus and minus one standard deviation is shown as the orange shaded region.  The shaded regions overlap only in the limit of very little pollution of the initial state.  The overlap region is consistent with no pollution and is inconsistent with pollution of the magnetization of greater than 5-10 atoms and pair pollution of greater than 3-5 pairs of $\pm1$ atoms, even for significantly different spinor dynamical rates than the evolution suggests.  While it is conceivable to trade off between spinor dynamical rate and pollution, the dynamics of the quadrature squeezing measurement reported previously \cite{Hamley12} indicates that the value of spinor dynamical rate estimated from the long term population dynamics is more consistent with the available data.  The analysis presented here along with the non-Gaussian nature of the fluctuations and the squeezing dynamics reported previously make an effective argument for the initial state preparation producing a very pure $m_f=0$ state.

\bibliography{QPref}
\bibliographystyle{unsrt}